\documentclass[aps,prb,twocolumn,amsmath,amssymb,nofootinbib,superscriptaddress,floatfix,longbibliography]{revtex4-2}

\usepackage{xcolor}
\usepackage{booktabs}
\usepackage{amsmath}
\usepackage{amssymb}
\usepackage{amsthm}
\usepackage{mathtools}
\usepackage{mathdots}
\usepackage{amsfonts}
\usepackage{bbm}
\usepackage{xr}
\usepackage{bbold}
\usepackage{epsfig,color,dsfont,upgreek,physics}
\usepackage{mathrsfs}
\usepackage{multirow}
\usepackage[makeroom]{cancel}
\usepackage{mathtools}

\usepackage{graphicx}
\usepackage{dcolumn} 
\usepackage{bm} 
\usepackage{float} 
\usepackage[driverfallback=dvipdfm]{hyperref} 
\usepackage{ulem}   
\hypersetup{
    colorlinks=true,
    linkcolor=blue,
    citecolor=blue,
    urlcolor=blue
}

\allowdisplaybreaks

\begin{document}

\title{Logarithmic Entanglement and Emergent Dipole Symmetry from a Strongly Coupled Light-Matter Quantum Circuit}

\author{Luiz H. Santos}
\affiliation
{
Department  of  Physics,  Emory  University,  400 Dowman Drive, Atlanta,  GA  30322,  USA
}

\begin{abstract}
Hybrid systems where a quantum material strongly couples to a nonlocal 
cavity photon mode have emerged as a new frontier for controlling and 
probing quantum correlations, yet the structure and scaling of 
light-matter entanglement produced by the nonlocal coupling remains 
poorly understood. We address this problem through an exactly solvable 
framework based on reinterpreting the Power--Zienau--Woolley (PZW) 
transformation as a \textit{light-matter quantum circuit} that couples 
the photonic position quadrature $X \sim a + a^\dagger$ to the many-body 
dipole $\mathcal{P}$ of a one-dimensional quantum chain. We derive a 
closed-form expression for the reduced density matrix valid at all 
coupling strengths, in which off-diagonal elements between matter states 
of unequal dipole are suppressed by a Gaussian factor encoding the full 
weak-to-ultrastrong coupling crossover. At weak coupling, the reduced 
density matrix takes a Lindbladian form with $\mathcal{P}$ as the jump 
operator, and the entanglement entropy is controlled by the dipole 
variance. At ultrastrong coupling, the density matrix becomes exactly 
block-diagonal in dipole sectors, reflecting an \textit{emergent dipole 
symmetry} dynamically imposed by the photon field, with entanglement 
entropy given exactly by the Shannon entropy of the dipole-sector weight 
distribution. Applying this framework to a half-filled 
Su--Schrieffer--Heeger chain, we show that, at strong coupling, both the light-matter 
entanglement and the spatial entanglement of the photon-dressed matter 
state scale logarithmically with system size, $S_\infty \sim 
\frac{\alpha}{2}\log L$, robust across the SSH phase diagram. The logarithm originates from the 
photon resolving a single collective coordinate $\mathcal{P}$ whose 
fluctuations grow as $L^{\alpha/2}$, a distinct mechanism from the 
logarithmic entanglement of critical one-dimensional systems. 
\end{abstract}
\date{\today}

\maketitle

\noindent

\section{Introduction}

Entanglement lies at the heart of quantum mechanics, providing correlations that have no
classical analog and underpinning both fundamental phenomena and emerging technologies.
In addition to being a key resource in quantum information protocols such as
teleportation~\cite{bennett1993teleporting, bouwmeester1997experimental,
pirandola2015advances, hu2023progress} and in prospective quantum computing
architectures~\cite{preskill2018quantum}, entanglement plays a central role in
characterizing quantum many-body systems, shaping topologically ordered
phases~\cite{kitaev-preskill-2006, levin-wen-2006, Li_Haldane_2008},  short-range entangled
symmetry-protected topological
states~\cite{affleck1988valence, pollmann2010entanglement, Fidkowski_Entanglement_2010, turner2011topological,
chen2011classification, schuch2011classifying, Santos_RK_2015, Santos_Symmetry-protected_2018}, critical
systems~\cite{holzhey1994geometric, calabrese2004entanglement, fradkin2006entanglement, bueno2015universality, refael2004entanglement},
holographic
dualities~\cite{ryu2006holographic}, and nontrivial dynamical regimes in nonergodic quantum
matter~\cite{bardarson2012unbounded, serbyn2013universal, serbyn2021quantum}.

\begin{figure}
\includegraphics[width=6.0cm]{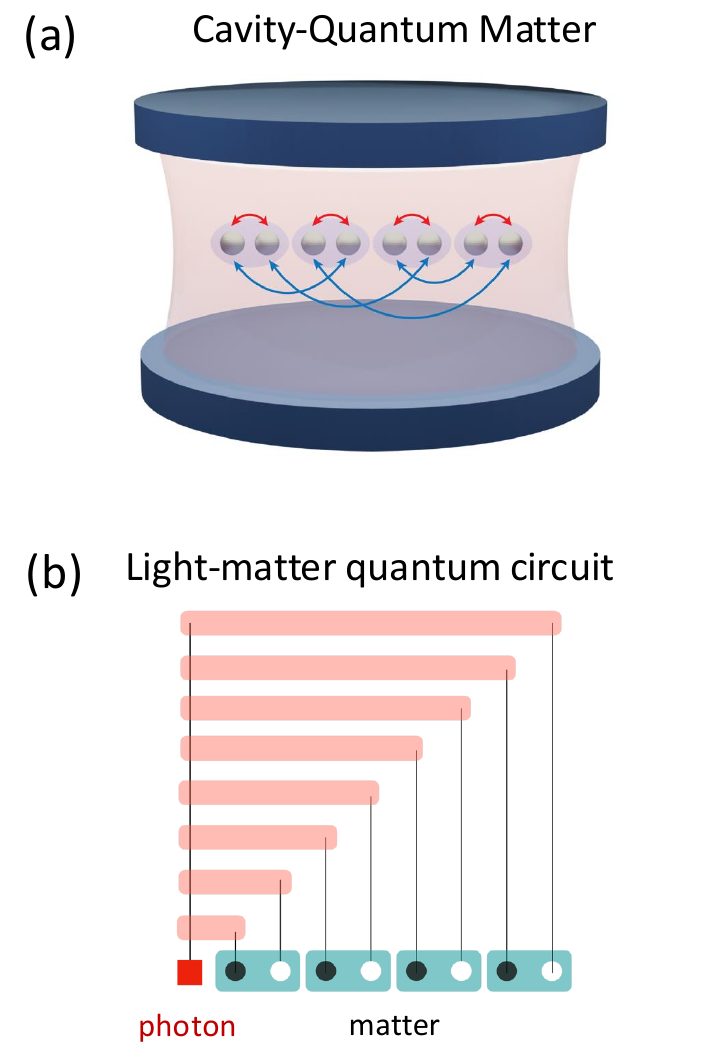}
\caption{(a) Cavity-quantum matter setup: a one-dimensional fermionic chain of length $L$
embedded in a single-mode optical cavity. (b) The Power--Zienau--Woolley (PZW)
transformation, reinterpreted as a quantum circuit that entangles the cavity photon with
all matter degrees of freedom by coupling the photonic position operator
$X \sim a + a^\dagger$ to the many-body dipole $\mathcal{P}$.}
\label{fig:cavity}
\end{figure}

The scaling of the entanglement entropy $S_A$ of a subsystem $A$ of linear dimension $\ell$
in $D$ spatial dimensions provides key insights into the nature of correlations between the
subsystem and its surroundings, informing the classification of states according to volume
law $S_A \sim \ell^D$ and area law $S_A \sim \ell^{D-1}$~\cite{eisert2010colloquium, hastings2007area}.  
In $D = 1$ critical systems
described by a conformal field theory of central charge $c$, the entanglement entropy
acquires a universal logarithmic scaling $S_A = \frac{c}{3}\log\ell$\cite{holzhey1994geometric, calabrese2004entanglement}.
These classifications
are, however, predicated on systems with local interactions, and it is natural to ask how they are modified when matter couples to a non-local degree of freedom. Recent experimental advances
have enabled strong light--matter interactions in cavity quantum materials, where an extended
material system couples to a subwavelength optical mode, as shown in Fig.~\ref{fig:cavity}(a).
The strong subwavelength mode confinement creates an effective long-range interaction
mediated by a spatially quasi-uniform photon mode in the strong-coupling
regime~\cite{Garcia-Vidal_Manipulating_2021, hubener2021engineering, FriskKockum2019,
Forn-Diaz_RMP_2019}, generating long-range photon-mediated correlations that fall
outside the local-interaction framework underlying standard entanglement
classifications~\cite{eisert2010colloquium, hastings2007area,
holzhey1994geometric, calabrese2004entanglement}.
This regime has attracted significant interest for its potential to
control and modify quantum material properties through coupling to quantum electromagnetic
fields confined in cavities, with theoretical and experimental studies demonstrating
cavity-induced modifications across a broad range of
platforms~\cite{Schlawin_Cavity-Mediated_2019, allocca2019cavity, Ashida_Quantum_2020,
latini2021ferroelectric, jarc2023cavity}. (For a recent review, see Ref. \onlinecite{bretscher2026fluctuationengineeringcavityquantum}.)

The entanglement properties of these hybridized light--matter eigenstates 
are a direct signature of the photon-mediated correlations generated by 
the cavity, yet their structure and scaling with system size remain 
poorly understood. These eigenstates take the general form
\begin{equation}
\label{eq: general LMI state}
\ket{\psi} =
\sum_{\alpha,\beta} \psi_{\alpha,\beta}
\ket{\textrm{matter}_{\alpha}} \otimes \ket{\textrm{light}_{\beta}}\,,
\end{equation}
and the associated light--matter entanglement is quantified by the von Neumann entropy
$S = -\mathrm{Tr}(\rho\log\rho)$, where
\begin{equation}
\rho = \sum_{\beta}\sum_{\alpha,\alpha'}
\psi_{\alpha,\beta}\,\psi^{*}_{\alpha',\beta}
\ket{\textrm{matter}_{\alpha}}\bra{\textrm{matter}_{\alpha'}}
\end{equation}
is the reduced density matrix obtained by tracing out the photon degrees of freedom.
Can the coupling of matter to a nonlocal cavity photon mode give rise to new regimes of
entanglement, both between light and matter, and within the matter sector itself?
And how does the entanglement depend on system size $L$ and on the light--matter coupling
strength, particularly when interactions become strong?

These questions are brought into focus by DMRG observations of weak, 
subextensive scaling of the ground-state entanglement entropy with system 
size in one-dimensional chains, rather than the extensive growth one might 
naively expect from the all-to-all photon-matter coupling~\cite{shaffer2024entanglement}. 
Related studies have investigated 
entanglement and topology in cavity-embedded systems from complementary 
perspectives, including topological markers and quantum transport in 
SSH-cavity chains~\cite{nguyen2024electron}, topological protection of 
Majorana polaritons~\cite{bacciconi2024majorana}, electron-photon Chern 
numbers in moir\'e materials~\cite{nguyen2023electronphoton}, and 
fractional quantum Hall liquids coupled to quantum light~\cite{bacciconi2025fqh}.
In these works, however, entanglement is either used as a diagnostic for 
the breakdown of mean-field theory~\cite{nguyen2024electron}, or 
characterized through entanglement spectra and electronic purity rather 
than through von Neumann entropy scaling~\cite{bacciconi2024majorana, 
nguyen2023electronphoton, bacciconi2025fqh}. The lack of a general 
understanding about the structure and scaling of entanglement in hybrid 
systems where photonic cavity modes interact with an extended material 
calls for analytically tractable models that can provide insights into 
the mechanism of entanglement in such systems.

In this work we introduce such a framework, characterizing the entanglement of a cavity photon mode interacting with a one-dimensional 
chain of length $L$. Our approach is based on the Power--Zienau--Woolley 
(PZW) unitary formulation of light--matter 
interactions~\cite{power1957radiative, power1959coulomb, 
woolley1971molecular}. While the PZW transformation is standard in 
quantum optics and cavity QED, we reinterpret it as a \textit{light-matter quantum circuit} 
-- shown in Fig.~\ref{fig:cavity}(b) -- that entangles the cavity photon 
with all matter degrees of freedom by coupling the photon position 
operator $X \sim a + a^\dagger$ to the many-body dipole $\mathcal{P}$. 
The hybrid quantum circuit, which is controlled by the dimensionless light-matter coupling $g$, acts on an unentangled product state of matter and light degrees of freedom, generating a family of photon--matter entangled states whose entanglement properties can be 
characterized \emph{exactly} at all coupling strengths.
\\

\noindent
\textit{Main results.---}
Central to our analysis is the reduced density matrix obtained by tracing out
the photon from the PZW-circuit state (Sec.~\ref{sec: Light--matter quantum circuit}).
We derive a closed-form expression revealing a simple but far-reaching structure: each
off-diagonal matrix element between matter Fock states $\ket{\{n\}}$ and $\ket{\{n'\}}$
is suppressed by a Gaussian factor
$e^{-\frac{g^2}{2}(\mathcal{P}(\{n\})-\mathcal{P}(\{n'\}))^2}$,
where $g$ is the dimensionless light--matter coupling and $\mathcal{P}(\{n\})$ is the
many-body dipole of the state $\ket{\{n\}}$. This exact result interpolates continuously
between two physically transparent limits.

At weak coupling $g \ll 1$, the reduced density matrix takes a Lindbladian
form~\cite{lindblad1976generators, gorini1976completely, 
breuer2002theory} with the dipole $\mathcal{P}$ as the jump operator and rate
$\gamma \propto g^2$ (Sec.~\ref{sec: Light--matter quantum circuit}). The Hermitian
nature of the jump operator reflects the fact that the photon acts as a monitor of the
many-body dipole: it acquires information about dipole fluctuations while leaving states
of definite dipole unaffected. As a result, the entanglement entropy is controlled
entirely by the dipole variance,
$S \approx \gamma\,\mathrm{Var}(\mathcal{P})\bigl[1 - \log(\gamma\,\mathrm{Var}(\mathcal{P}))\bigr]$,
and grows quasi-linearly with system size for a half-filled chain.

At ultrastrong coupling $g \gg 1$, the Gaussian suppression is complete: coherence 
survives only between matter configurations sharing the same many-body dipole moment, 
and the reduced density matrix becomes exactly block-diagonal in dipole sectors 
(Sec.~\ref{sec: Scaling Analysis of the Entanglement Entropy}). 
This block-diagonal structure reflects an \emph{emergent dipole symmetry} dynamically imposed by the photon field.
At this regime, the entanglement entropy reduces \textit{exactly} to the Shannon entropy 
of the dipole-sector weight distribution, $S_\infty = -\sum_{\mathcal{P}} 
p(\mathcal{P})\log p(\mathcal{P})$. Importantly, comparisons between numerical simulations and the exact reduced density matrix at the strong coupling reveal that 
the emergent dipole symmetry and the associated logarithmic entanglement scaling are well established at 
$g \sim \mathrm{O}(1)$ coupling strengths. 
The suppression of single-particle hopping and the survival of only dipole-conserving 
processes in this limit bears a direct structural resemblance to the dipole-conserving 
Hamiltonians that underlie Hilbert space 
fragmentation~\cite{sala2020ergodicity, khemani2020localization, moudgalya2022review}, 
though here the dipole constraint is not intrinsic to the matter Hamiltonian but is 
dynamically generated by the cavity photon.

The key results of Sec.~\ref{sec: Scaling Analysis of the Entanglement Entropy} establish
that both the light--matter entanglement and the spatial entanglement of the photon-dressed
matter state scale logarithmically with system size as the coupling increases.
In particular, we present our analysis for a half-filled
Su--Schrieffer--Heeger (SSH) chain  with intra- and inter-unit-cell hopping
$t_1$ and $t_2$ coupled to a single cavity photon. We establish that the light--matter entanglement entropy and the half-chain matter
entanglement of the photon-dressed state satisfy, respectively,
\begin{equation*}
S_\infty \sim \tfrac{\alpha}{2}\log L \,,
\qquad
S^{\mathrm{matter}}_\infty \propto \log\ell\,,
\end{equation*}
where $\ell$ is the subsystem size. Both scalings share the same mechanism: the photon
resolves only the single collective coordinate $\mathcal{P}$, whose fluctuations grow as
$L^{\alpha/2}$, so the entanglement is controlled by the statistics of this variable
rather than by the full many-body Hilbert space. In the analytically tractable dimerized
limit $t_2 = 0$, the dipole distribution is exactly Gaussian with $\alpha = 1$, giving
$S_\infty \approx \frac{1}{2}\log L$. Crucially, both scalings arise in states that
are topologically trivial --- obtained by a unitary transformation from a gapped,
disentangled product state --- and are therefore distinct in origin from the logarithmic
entanglement of critical one-dimensional
systems~\cite{holzhey1994geometric, calabrese2004entanglement}: the logarithm here is a
consequence of non-local photon-mediated coupling rather than conformal invariance. The
prefactor $\alpha/2$ is not universal in the CFT sense: it evolves continuously with $g$,
saturating at its ultrastrong-coupling value, and grows with the correlation length of the
SSH chain, reaching its maximum at the critical point $t_1 = t_2$.

More broadly, our results show that coupling an extended quantum system to a confined 
electromagnetic mode generates a qualitatively distinct entanglement structure -- one 
controlled not by the full many-body Hilbert space but by the statistics of a single 
collective coordinate selected by the photon. This reduction of complexity is a direct 
consequence of the non-local, mode-uniform nature of the cavity coupling, and it 
suggests that the entanglement properties of cavity-matter systems are governed by 
different organizational principles than those of locally interacting systems. 
Characterizing these principles -- how they depend on the geometry of the photon mode, 
the dimensionality of the matter system, and the number of cavity modes -- may reveal 
new organizational patterns in hybrid light-matter systems more broadly, 
and provide new handles for controlling quantum correlations through electromagnetic 
confinement.

\section{Light--matter quantum circuit}
\label{sec: Light--matter quantum circuit}

We consider a finite one-dimensional chain with $L = 2N$ sites populated by spinless fermions.
The fermionic degrees of freedom are described by creation and annihilation operators $c^{\dagger}_j$ and $c_j$.
The matter sector is governed by the free-fermion Hamiltonian
\begin{equation}
H_{\textrm{m}} = -\sum_{j}
\left(
t_{j} c^{\dagger}_{j} c_{j+1} + \textrm{H.c.}
\right),
\end{equation}
where $t_j$ denotes the nearest-neighbor hopping amplitude.
For concreteness we restrict to nearest-neighbor hopping, though the results below generalize straightforwardly to longer-range processes.

Since $H_{\textrm{m}}$ conserves particle number, it block-diagonalizes within sectors of fixed fermion number.
We focus on the half-filled sector spanned by the Fock basis
$\{ \ket{n_1, \ldots, n_L} \mid \sum^{L}_{j=1} n_{j} = N \}$.
The eigenstates $\ket{\Phi_{\alpha}}$ of $H_{\textrm{m}}$, defined by
$H_{\textrm{m}} \ket{\Phi_{\alpha}} = \varepsilon_{\alpha} \ket{\Phi_{\alpha}}$,
expand in this basis as
\begin{equation}
\ket{\Phi_{\alpha}}
=
\sum_{\{n\}}
\Phi_{\alpha}(\{n\})
\ket{n_1, \ldots, n_L}.
\end{equation}

The fermionic chain, oriented along the $x$ axis, is embedded in an optical cavity supporting a
transverse electromagnetic field described by the vector potential
\begin{equation}
\mathbf{A} = \mathcal{A}_{0} \left( a + a^{\dagger} \right) \hat{e}_{x},
\end{equation}
where $a$ and $a^{\dagger}$ are bosonic operators satisfying $[a,a^{\dagger}] = 1$,
and the field amplitude $\mathcal{A}_{0} = \sqrt{\hbar / (2 \omega V \epsilon)}$,
with $V$ the cavity volume and $\epsilon$ the dielectric constant.
The photon mode is governed by
\begin{equation}
H_{\textrm{ph}} = \hbar \omega \, a^{\dagger} a,
\end{equation}
whose eigenstates satisfy $H_{\textrm{ph}} \ket{m} = \hbar \omega m \ket{m}$, with $m = 0,1,\ldots$.

In the absence of light--matter coupling, the total Hamiltonian
\begin{equation}
H_{0} = H_{\textrm{m}} + H_{\textrm{ph}}
\end{equation}
describes fully decoupled matter and photon sectors, with product eigenstates
\begin{equation}
\begin{split}
H_{0} \ket{\Phi_{\alpha,m}} &= (\varepsilon_{\alpha} + m \hbar\omega) \ket{\Phi_{\alpha,m}},
\\
\ket{\Phi_{\alpha,m}} &= \ket{\Phi_{\alpha}} \otimes \ket{m}.
\end{split}
\end{equation}

Light--matter interactions are incorporated via a Peierls substitution, yielding
\begin{equation}
\label{eq: H cavity-matter Peierls}
H(g) =
-\sum_{j=1}^{L-1}
\Big(
t_{j} \,
e^{i g (a + a^{\dagger})}
c^{\dagger}_{j} c_{j+1}
+ \textrm{H.c.}
\Big)
+
\hbar \omega a^{\dagger} a,
\end{equation}
where the dimensionless coupling $g \equiv e \ell \mathcal{A}_{0} / \hbar$,
with $\ell$ the lattice spacing, controls the strength of the light--matter interaction.

A key observation is that the Hamiltonian~\eqref{eq: H cavity-matter Peierls} can be written as
\begin{equation}
\label{eq: LMI H expressed via PZW}
H(g) = \mathcal{U}(g) \, H_{\textrm{m}} \, \mathcal{U}(g)^{-1}
+ \hbar \omega a^{\dagger} a,
\end{equation}
where
\begin{equation}
\mathcal{U}(g) = e^{-i g (a + a^{\dagger}) \mathcal{P}}
\end{equation}
is the Power--Zienau--Woolley (PZW) unitary transformation~\cite{power1957radiative, power1959coulomb, woolley1971molecular}, and
\begin{equation}
\mathcal{P} = \sum_{j=1}^{L}
(j - \tilde{x}) \, c^{\dagger}_{j} c_{j}
\end{equation}
is the dimensionless many-body dipole operator, with $\tilde{x} = \frac{L+1}{2}$ the midpoint of the chain.%
\footnote{Under a global shift of coordinates $j \rightarrow j + \tilde{x}$,
the polarization transforms as $\mathcal{P} \rightarrow \mathcal{P} + \tilde{x} \hat{N}$,
which is inconsequential since particle number is conserved.}

Under $\mathcal{U}(g)$, the fermionic and photonic operators transform as
\begin{equation}
\begin{split}
\mathcal{U}(g)
\,
c_j
\,
\mathcal{U}(g)^{-1}
&=
e^{i g (a + a^{\dagger}) (j - \tilde{x})} \, c_j,
\\
\mathcal{U}(g)\, a \, \mathcal{U}(g)^{-1}
&=
a + i g \, \mathcal{P},
\\
\mathcal{U}(g) \, a^{\dagger} \, \mathcal{U}(g)^{-1}
&=
a^{\dagger} - i g \, \mathcal{P}.
\end{split}
\end{equation}

This structure motivates reinterpreting the PZW transformation as a quantum circuit that entangles
a single photon with all matter degrees of freedom by coupling the photonic position operator
$X \sim a + a^{\dagger}$ to the many-body dipole $\mathcal{P}$,
as illustrated in Fig.~\ref{fig:cavity}.
Acting on an unentangled product state, this circuit generates the family of entangled states
\begin{equation}
\label{eq: general state as fcn of m}
\ket{{\Psi}_{\alpha,m}(g)}
\equiv
\mathcal{U}(g)
\ket{\Phi_{\alpha}} \otimes \ket{m}.
\end{equation}
At $g=0$ the state~\eqref{eq: general state as fcn of m} is a product state,
while at finite coupling the PZW circuit generates entanglement between light and matter.
We emphasize that the states $|\Psi_{\alpha,m}(g)\rangle$ are eigenstates of the 
circuit Hamiltonian
\begin{equation}
\label{eq: H circuit}
H_{\mathcal{U}} = \mathcal{U}(g)\left(H_{\textrm{m}} + \hbar\omega a^{\dagger}a\right)\mathcal{U}(g)^{-1},
\end{equation}
rather than of $H(g)$ itself. Using the explicit form of the PZW 
transformation~\eqref{eq: LMI H expressed via PZW}, one can show that
\begin{equation}
\label{eq: H circuit vs H}
H_{\mathcal{U}} = H(g) + \Delta(g),
\end{equation}
where
\begin{equation}
\label{eq: Delta}
\Delta(g) = \hbar\omega\left[g\,i(a^{\dagger} - a)\mathcal{P} + g^2\mathcal{P}^2\right].
\end{equation}
The correction $\Delta(g)$ vanishes at $g = 0$ and grows with both the coupling 
strength and the many-body dipole $\mathcal{P}$. 
At weak coupling $g \ll 1$, 
$H_{\mathcal{U}}$ provides a controlled approximation to $H(g)$, while at stronger 
coupling the two Hamiltonians differ significantly by the action of $\mathcal{P}$-dependent contributions.
The entanglement analysis that 
follows characterizes this exactly solvable family of light--matter states as a 
tractable framework for understanding photon-mediated correlations in cavity quantum 
systems. This family of states provides the basis for our analytical and numerical 
characterization of the weak-to-strong coupling crossover carried out below.

We begin the entanglement analysis by specializing to the case where the circuit acts on
a generic matter state $\ket{\Phi}$ with the photon vacuum,
\begin{equation}
\label{eq: entangled state from |0> photon state}
\ket{{\Psi}_{0}(g)}
=
\mathcal{U}(g)
\ket{\Phi} \otimes \ket{0},
\end{equation}
where the matter eigenstate index $\alpha$ is suppressed,
as the analysis does not rely on any specific structure of $\ket{\Phi}$.

Expressing the photon operators in terms of the oscillator position and momentum,
$a = \sqrt{\omega/(2\hbar)} (x + i p/\omega)$,
the PZW transformation takes the form
\begin{equation}
\mathcal{U}(g)
=
e^{-i g \sqrt{2\omega/\hbar} \, x \, \mathcal{P}}.
\end{equation}
Writing the matter state as
\begin{equation}
\ket{\Phi} = \sum_{\{ n \}} \Phi(\{ n \}) \ket{\{ n \}}
\end{equation}
and the photon vacuum as
\begin{equation}
\ket{0}
=
\int dx \, \xi_{0}(x) \ket{x},
\end{equation}
with $\xi_{0}(x) = \pi^{-1/4} \sigma^{-1/2} e^{-x^{2}/(2\sigma^{2})}$ and $\sigma = (\hbar/\omega)^{1/2}$,
we obtain
\begin{equation}
\ket{{\Psi}_{0}(g)}
=
\int dx
\sum_{\{ n \}}
\Phi(\{ n \})\,\xi_{0}(x)\,
e^{-i g \sqrt{\frac{2\omega}{\hbar}}\, x\, \mathcal{P}(\{ n \})}
\ket{\{ n \}} \otimes \ket{x}.
\end{equation}

The light--matter entanglement encoded in the PZW circuit is extracted from the reduced density
matrix obtained by tracing out the photon,
\begin{equation}
\rho_{0}
=
\int dx\, \langle x \ket{{\Psi}_{0}(g)} \bra{{\Psi}_{0}(g)} x\rangle.
\end{equation}
The Gaussian nature of the photon vacuum and the linearity of the coupling between
$\mathcal{P}$ and the photon coordinate $x$ allow this integral to be performed exactly,
yielding the closed-form reduced density matrix
\begin{equation}
\label{eq: photon DM state 0}
\begin{split}
\rho_{0}
&\,=
\sum_{\{ n \},\{ n' \}}
\Phi(\{ n \}) \Phi^{*}(\{ n' \})\,
e^{-\frac{g^{2}}{2} \left( \mathcal{P}(\{ n \}) - \mathcal{P}(\{ n' \}) \right)^{2}}
\\
&\,\qquad\qquad\qquad
\times
\ket{\{ n \}} \bra{\{ n' \}},
\end{split}
\end{equation}
which is properly normalized, $\mathrm{Tr}(\rho_{0}) = 1$.

Equation~\eqref{eq: photon DM state 0} generalizes straightforwardly to an arbitrary initial
photon number $m$. Tracing out the photon from the state~\eqref{eq: general state as fcn of m} yields
\begin{equation}
\label{eq: photon DM state m}
\begin{split}
\rho_{m}
&\,=
\sum_{\{n\},\{n'\}}
\Phi(\{n\}) \Phi^{*}(\{n'\})\,
L_{m}\!\left(
g^{2} \left( \mathcal{P}(\{n\}) - \mathcal{P}(\{ n' \}) \right)^{2}
\right)
\\
&\,\qquad\qquad
\times
e^{-\frac{g^{2}}{2} \left( \mathcal{P}(\{ n \}) - \mathcal{P}(\{ n' \}) \right)^{2}}
\ket{\{ n \}} \bra{ \{ n' \}},
\end{split}
\end{equation}
where $L_{m}(x)$ denotes the Laguerre polynomial.
Expectation values of matter operators $\mathcal{O}_{\textrm{matter}}$ follow directly
from Eq.~\eqref{eq: photon DM state m},
\begin{equation}
\label{eq: matter operator expectation value}
\begin{split}
\langle
&\,\mathcal{O}_{\textrm{matter}}
\rangle
=
\mathrm{Tr} \Big( \rho_{m} \, \mathcal{O}_{\textrm{matter}}
\Big)
\\
&\,=\sum_{\{n\},\{n'\}}
\Phi(\{n\}) \Phi^{*}(\{n'\})\,
L_{m}\!\left(
g^{2} \left( \mathcal{P}(\{n\}) - \mathcal{P}(\{ n' \}) \right)^{2}
\right)
\\
&\,\qquad\qquad
\times
e^{-\frac{g^{2}}{2} \left( \mathcal{P}(\{ n \}) - \mathcal{P}(\{ n' \}) \right)^{2}}
\bra{ \{ n' \}}
\mathcal{O}_{\textrm{matter}}
\ket{\{ n \}}.
\end{split}
\end{equation}
At $g=0$, Eq.~\eqref{eq: photon DM state m} reduces to the pure state
$\rho(0) = \ket{\Phi}\bra{\Phi}$ for all $m$, and
$\langle \mathcal{O}_{\textrm{matter}} \rangle_{g=0}
= \bra{\Phi} \mathcal{O}_{\textrm{matter}} \ket{\Phi}$, as expected.

Equations~\eqref{eq: photon DM state 0} and~\eqref{eq: photon DM state m} are the central
results of this section. The reduced density matrices provide both qualitative and quantitative
access to the light--matter entanglement generated by the PZW circuit, which is quantified
by the von Neumann entropy
\begin{equation}
\label{eq: EE general}
S = - \mathrm{Tr}\Big( \rho_{m}\,\log{\rho_{m}} \Big).
\end{equation}
The remainder of this section develops a qualitative picture of the entanglement across
coupling regimes; a detailed quantitative analysis for the dimerized lattice is deferred
to Sec.~\ref{sec: Scaling Analysis of the Entanglement Entropy}.

\medskip
\noindent\textit{Weak coupling.---}
At weak coupling $g \ll 1$, we expand the reduced density matrix~\eqref{eq: photon DM state m}
in powers of $g$ about the pure state $\rho(0) = \ket{\Phi}\bra{\Phi}$.
To leading order,
\begin{equation}
\label{eq: lowest order photon DM for m photons}
\delta \rho_{m}
=
(2m+1) g^{2}
\Big(
\mathcal{P} \rho(0) \mathcal{P}
-
\frac{1}{2} \{ \mathcal{P}^{2}, \rho(0) \}
\Big),
\end{equation}
where $\delta \rho_{m} \equiv \rho_{m}(g) - \rho(0)$.
Equation~\eqref{eq: lowest order photon DM for m photons} has the form of a
Lindbladian~\cite{breuer2002theory} with Hermitian jump operator $\mathcal{P}$ and
rate $\gamma_{m} = (2m+1)g^{2}$, which grows with photon occupation.
This structure reveals that the photon monitors the many-body dipole, inducing decoherence
between states with different dipole moments while leaving states of definite dipole
unaffected --- the right-hand side of Eq.~\eqref{eq: lowest order photon DM for m photons}
vanishes identically when $\rho(0)$ has definite dipole.

To quantify the entanglement generated at weak coupling, we track the leading-order shift
of the single nonzero Schmidt eigenvalue, $p \rightarrow 1 - \delta$.
A direct calculation using Eq.~\eqref{eq: lowest order photon DM for m photons} gives
\begin{equation}
\begin{split}
\delta
&\,=
-\bra{\Phi} \delta \rho_m \ket{\Phi}
\\
&\,=
-(2m+1)g^2
\Big[
\bra{\Phi} \mathcal{P} \rho(0) \mathcal{P} \ket{\Phi}
-
\bra{\Phi}
\tfrac{1}{2} \{ \mathcal{P}^{2}, \rho(0)\} \ket{\Phi}
\Big]
\\
&\,=
(2m+1)g^2
\Big[
\bra{\Phi} \mathcal{P}^2 \ket{\Phi}
-
\bra{\Phi} \mathcal{P} \ket{\Phi}^2
\Big]
\\
&\,=
(2m+1)g^2\,\mathrm{Var}(\mathcal{P}),
\end{split}
\end{equation}
from which the light--matter entanglement entropy at weak coupling follows as
\begin{equation}
\label{eq: light-matter entanglement at weak coupling}
\begin{split}
S_{g \ll 1}
&\,\approx
-(1 - \delta)\log{(1-\delta)} - \delta \log{\delta}
\\
&\,\approx
(2m+1)g^2\,\mathrm{Var}(\mathcal{P})
\Big[
1 - \log\!\Big( (2m+1)g^2\,\mathrm{Var}(\mathcal{P})
\Big)
\Big].
\end{split}
\end{equation}
The dependence of Eq.~\eqref{eq: light-matter entanglement at weak coupling} on the dipole
variance reflects the role of the photon as a monitor of the many-body dipole: the photon
acquires information about dipole fluctuations, while states of definite dipole are annihilated
by the right-hand side of Eq.~\eqref{eq: lowest order photon DM for m photons}.
Consequently, the variance of the dipole operator controls both the scale of the light--matter
entanglement and its system-size scaling, which we analyze in
Sec.~\ref{sec: Scaling Analysis of the Entanglement Entropy}.

\medskip
\noindent\textit{Ultrastrong coupling.---}
As the coupling increases, the Gaussian factor in Eq.~\eqref{eq: photon DM state m}
progressively suppresses coherence between matter states with different dipole moments.
In the ultrastrong coupling limit $g \gg 1$, this suppression becomes complete: all
off-diagonal matrix elements connecting sectors of unequal dipole vanish, and the reduced
density matrix retains only blocks satisfying
$\mathcal{P}(\{n\}) = \mathcal{P}(\{n'\})$.
The reduced density matrix~\eqref{eq: photon DM state m} therefore tends asymptotically to
\begin{equation}
\label{eq: RDM at ultra-strong coupling}
\rho_{\infty}
=
\sum_{\{ n \},\{ n' \}}
\Phi(n) \Phi^{*}(n')\,
\delta_{\mathcal{P}(\{n\}), \mathcal{P}(\{n'\})}
\ket{\{n\}} \bra{\{n'\}},
\end{equation}
independently of the photon number $m$.

Defining $S(\mathcal{P}) = \{ \ket{\{n \}} \mid \sum_{j} j \, n_{j} = \mathcal{P}\}$
as the subspace of Fock states with fixed dipole $\mathcal{P}$, the asymptotic density
matrix~\eqref{eq: RDM at ultra-strong coupling} takes the block-diagonal form
\begin{equation}
\label{eq: RDM direct sum}
\rho_{\infty}
=
\bigoplus_{\mathcal{P}}
p({\mathcal{P}}) \, \rho({\mathcal{P}}),
\end{equation}
where $\rho({\mathcal{P}}) = \ket{\Phi({\mathcal{P}})}\bra{\Phi({\mathcal{P}})}$
is the normalized density matrix within the fixed-dipole subspace,
\begin{equation}
\ket{\Phi({\mathcal{P}})}
=
p({\mathcal{P})}^{-1/2}
\sum_{\{ n \} \in S({\mathcal{P}})}
\Phi(\{ n \}) \ket{\{ n \}},
\end{equation}
and
\begin{equation}
p(\mathcal{P}) = \sum_{\{ n \} \in S(\mathcal{P})} |\Phi(\{n\})|^{2}
\end{equation}
is the total probability weight of that sector.
The block-diagonal structure of Eq.~\eqref{eq: RDM direct sum} reflects an emergent
dipole symmetry at ultrastrong coupling, and the corresponding entanglement entropy
takes the form
\begin{equation}
\label{eq: dipole expression for EE}
S_{\infty}
=
- \sum_{\mathcal{P}} p(\mathcal{P}) \log p(\mathcal{P}).
\end{equation}

As shown in Sec.~\ref{sec: Scaling Analysis of the Entanglement Entropy}, the decomposition
of the Hilbert space into emergent dipole sectors gives rise to the scaling law
\begin{equation}
S_{\infty} \propto \log L.
\end{equation}
This subextensive scaling reflects a sharp distinction from conventional Markovian baths:
rather than inducing uniform decoherence across the full Hilbert space, the cavity photon
selectively suppresses coherence only between sectors of different dipole moment,
leaving the intra-sector structure entirely intact.

We close this section by noting an important consequence of the ultrastrong coupling limit
for matter observables. From Eq.~\eqref{eq: RDM at ultra-strong coupling}, the expectation
value of any matter operator takes the form
\begin{equation}
\label{eq: matter operator expectation value - strong coupling}
\begin{split}
&\,\langle \mathcal{O}_{\textrm{matter}} \rangle_{\infty} = 
\\
&\,
\sum_{\{ n \},\{ n' \}}
\Phi(n) \Phi^{*}(n')\,
\delta_{\mathcal{P}(\{n\}), \mathcal{P}(\{n'\})}
\bra{\{n'\}}
\mathcal{O}_{\textrm{matter}}
\ket{\{n\}}.
\end{split}
\end{equation}
Operators that do not conserve the dipole moment therefore acquire vanishing expectation
values, confirming the emergent dipole symmetry. In particular, single-particle hopping
$c^{\dagger}_{i} c_{j}$ is suppressed, while dipole-conserving correlated processes such
as $c^{\dagger}_{j} c_{j+m} c^{\dagger}_{i} c_{i-m}$ remain active --- describing
density-dependent interactions ($m=0$) or correlated hopping ($m \neq 0$). This structure
is directly analogous to the dipole-conserving Hamiltonians studied in the context of
Hilbert space fragmentation~\cite{sala2020ergodicity, khemani2020localization,
moudgalya2022review}, where conservation of the dipole moment under Hamiltonian dynamics
partitions the Hilbert space into dynamically disconnected sectors. A key distinction,
however, is that in our model the dipole symmetry is not an intrinsic property of the
matter Hamiltonian but is instead dynamically imposed by the photon field in the
ultrastrong coupling limit, providing a new mechanism by which cavity confinement can
induce constrained dynamics in an otherwise unconstrained system.

\section{Scaling Analysis of the Entanglement Entropy}
\label{sec: Scaling Analysis of the Entanglement Entropy}

In this section we characterize the light-matter entanglement entropy $S(g,L)$ as a function
of both the coupling $g$ and the system size $L$. We trace the evolution from the weak-coupling
regime, governed by the perturbative density matrix Eq.~\eqref{eq: lowest order photon DM for m photons},
to the ultrastrong coupling regime, whose entanglement entropy is given by
$S_{\infty}(L)$ in Eq.~\eqref{eq: dipole expression for EE}.

\subsection{Light-matter entanglement}
\label{sec: Light-matter entanglement}

To gain intuition and analytical understanding, we begin with the exactly solvable limit $t_2 = 0$,
corresponding to a trivial gapped chain whose unique ground state is a product of decoupled dimers,
each occupied by a single electron,
\begin{equation}
\label{eq: dimer state}
\ket{\Phi_{\textrm{dimer}}} =
\prod^{L/2}_{j=1}
\frac{1}{\sqrt{2}}
\left(
\ket{0_{2j-1}}\otimes\ket{1_{2j}}
+
\ket{1_{2j-1}}\otimes\ket{0_{2j}}
\right).
\end{equation}
The states $\ket{1_{2j-1}}\otimes\ket{0_{2j}}$ and $\ket{0_{2j-1}}\otimes\ket{1_{2j}}$ correspond,
respectively, to the particle occupying the left and right sites of the dimer. We refer to the
former as the ``reference'' dimer state and the latter as a ``broken'' dimer configuration, a terminology that conveniently labels the space of dipole configurations.

\begin{figure}
\includegraphics[width=9.0cm]{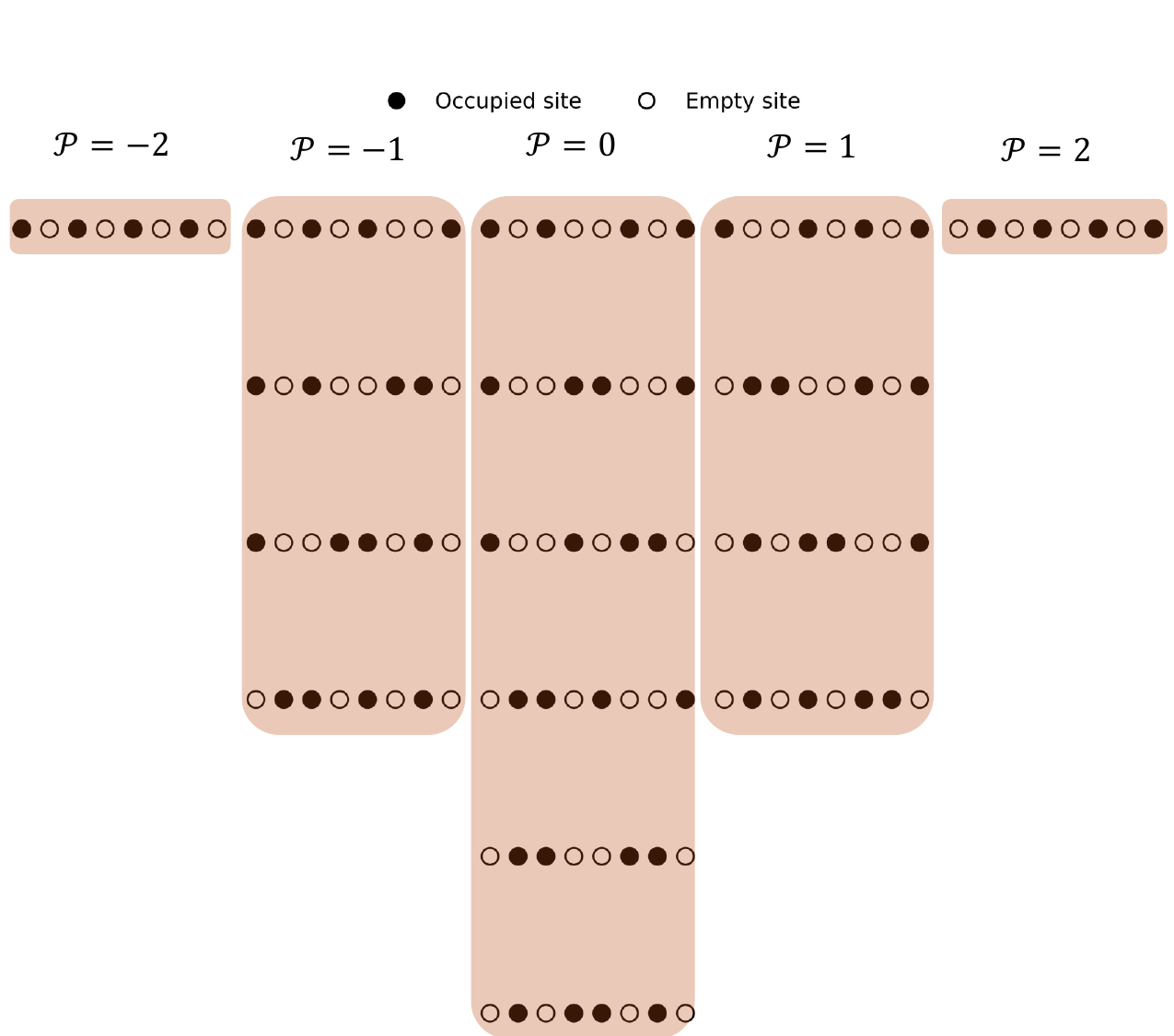}
\caption{Dipole-sector structure of the dimer state for $L = 8$, $N = 4$.
States are grouped by dipole value $\mathcal{P}_k = k - N/2$, $k = 0,\ldots,N$.
The binomial weight of each sector follows from choosing $k$ broken dimers
(right-site occupied) out of $N = L/2$ total dimers, forming the Pascal triangle shown.}
\label{fig: P configurations}
\end{figure}

With the dipole defined as $\mathcal{P} = \sum^{L}_{j=1}(j-\tilde{x})c^{\dagger}_{j}c_{j}$,
where $\tilde{x} = \frac{L+1}{2}$ is the chain midpoint, the state~\eqref{eq: dimer state} is
spanned by Fock states grouped into sectors of fixed dipole
\begin{equation}
\label{eq: broken dimers dipole relation}
\mathcal{P}_k = k - \frac{(L/2)}{2}\,, \quad k = 0, \ldots, L/2\,,
\end{equation}
where $k$ counts the number of broken dimers, i.e., those with their right site occupied.
The minimum dipole sector ($k = 0$) has all electrons on the left sites, while the maximum
($k = L/2$) has all electrons on the right sites. The probability weight of each dipole sector
is the binomial distribution over broken-dimer configurations,
\begin{equation}
p(\mathcal{P}_k) = 2^{-L/2}\,
\begin{pmatrix} L/2 \\ k \end{pmatrix},
\quad k = 0, \ldots, L/2\,,
\end{equation}
where $\binom{L/2}{k}$ counts the configurations with $k$ broken dimers.
This structure is illustrated in Fig.~\ref{fig: P configurations} for $L = 8$ and $N = 4$,
where grouping states by dipole value reveals the Pascal triangle of sector weights.

In the large $L$ limit, applying the Stirling approximation to the binomial and using
Eq.~\eqref{eq: broken dimers dipole relation} to convert from $k$ to $\mathcal{P}_k$, the
sector weights take a Gaussian form,
\begin{subequations}
\begin{equation}
\label{eq: Schmidt eigenvalues dimer state exponential form}
p(\mathcal{P}_k) \approx
\left(\frac{L\,\pi}{4}\right)^{-1/2}
e^{-\frac{\mathcal{P}_{k}^{2}}{(L/4)}}\,,
\end{equation}
\begin{equation}
\int^{\infty}_{-\infty}
dk \, p(\mathcal{P}_k)
=
\int^{\infty}_{-\infty} d\mathcal{P}\,
\left(\frac{L\,\pi}{4}\right)^{-1/2}
e^{-\frac{\mathcal{P}^{2}}{L/4}}\,.
\end{equation}
\end{subequations}
Thus the dimer state~\eqref{eq: dimer state} is characterized by a Gaussian distribution of
dipole moments centered at zero with variance $\sigma^2 = \mathrm{Var}(\mathcal{P}) = L/8 = N/4$.

\begin{figure}
\includegraphics[width=9.0cm]{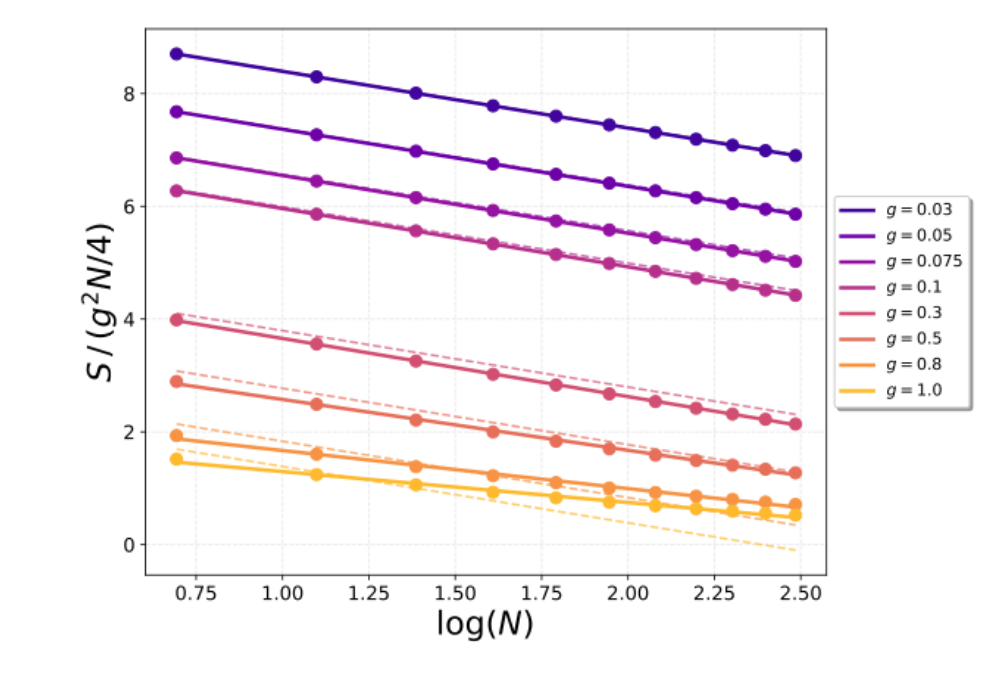}
\caption{Light-matter entanglement entropy at weak coupling.
Solid lines: exact entropy for the SSH ground state at $t_1 = 1$, $t_2 = 0$.
Dashed lines: quasi-linear prediction~\eqref{eq: S weak coupling}.
The two agree closely at small $g$; pronounced deviations develop at intermediate
to strong coupling as higher dipole sectors become resolved.}
\label{fig: Entanglement Weak Coupling}
\end{figure}

\noindent\textit{Weak coupling.---}
Substituting $\mathrm{Var}(\mathcal{P}) = N/4$ into the weak-coupling
result~\eqref{eq: light-matter entanglement at weak coupling} (restricting to $m = 0$ for
concreteness) gives
\begin{equation}
\label{eq: S weak coupling}
S \approx \frac{g^2\,N}{4}
\left[
1 - \log\!\left(\frac{g^2\,N}{4}\right)
\right],
\quad g^2 N \ll 1\,.
\end{equation}
At weak coupling the entanglement entropy grows quasi-linearly with $N$, controlled entirely
by the dipole variance. Figure~\ref{fig: Entanglement Weak Coupling} shows $S/N$ as a function
of $\log N$ computed from the general expression \eqref{eq: photon DM state 0} for several values of $g$: the quasi-linear prediction~\eqref{eq: S weak coupling}
agrees well with the exact numerical results at small $g$, while clear deviations develop as
the coupling increases.

\begin{figure*}[t]
\centering
\includegraphics[width=\textwidth]{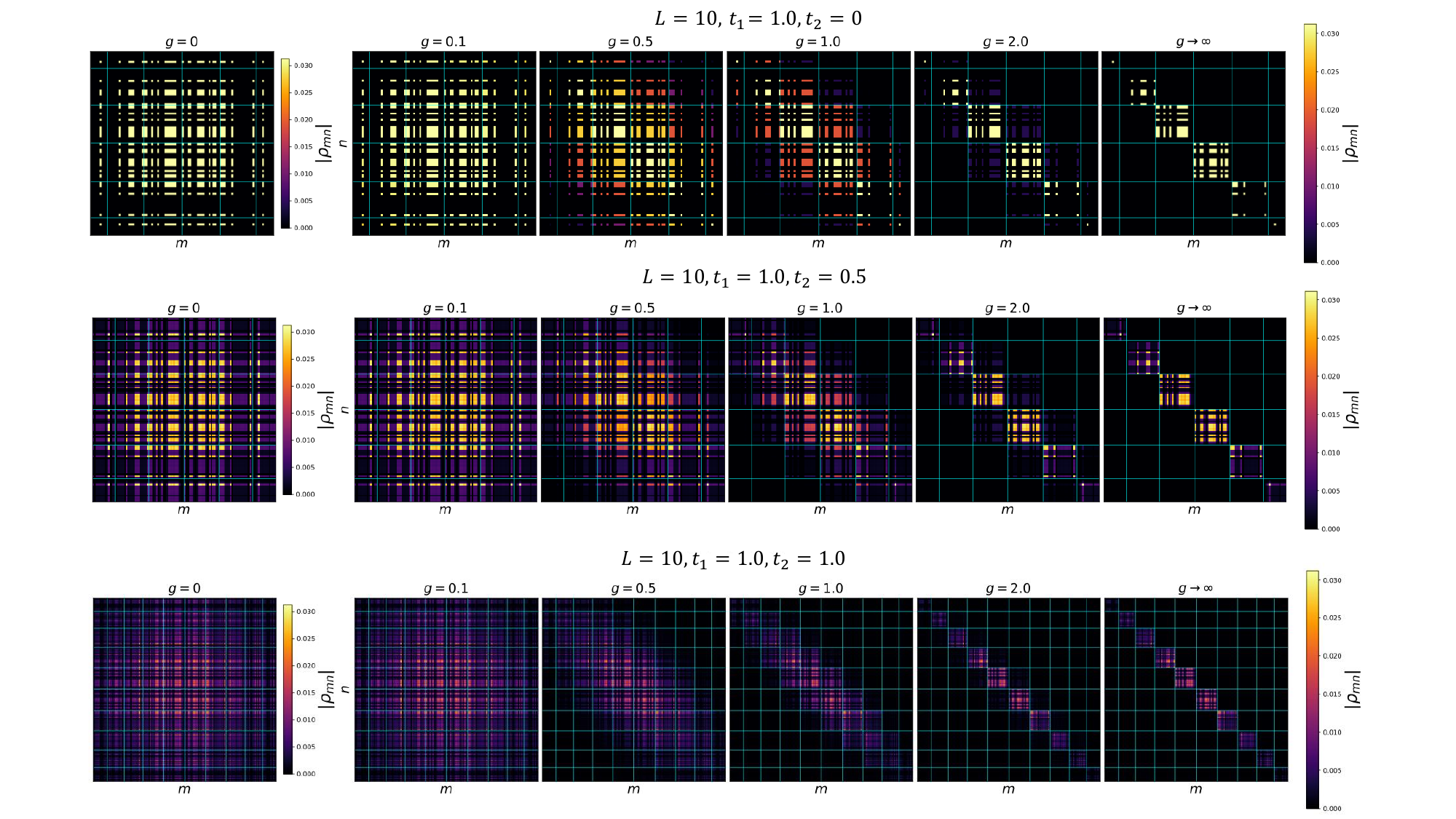}
\caption{Evolution of the reduced density matrix~\eqref{eq: photon DM state 0} for
$\ket{\Psi_0(g)} = \mathcal{U}(g)\ket{\Phi}\otimes\ket{0}$, where $\ket{\Phi}$ is the SSH
ground state on an $L = 10$ chain, shown as a function of $g$ for the completely dimerized
chain $(t_1,t_2) = (1,0)$ (top), intermediately dimerized case $(t_1,t_2) = (1,0.5)$ (middle),
and critical chain $(t_1,t_2) = (1,1)$ (bottom). In all cases the density matrix clusters into
the block-diagonal dipole-sector structure~\eqref{eq: RDM direct sum}, with well-formed sectors
by $g \sim 2$.}
\label{fig: evolution of density matrix}
\end{figure*}

\noindent\textit{Crossover to strong coupling.---}
As $g$ increases beyond the quasi-linear regime, the Gaussian suppression factor in the
reduced density matrix~\eqref{eq: photon DM state 0} progressively clusters coherences into
dipole sectors. This is directly visible in Fig.~\ref{fig: evolution of density matrix}, which
shows the evolution of the reduced density matrix for the SSH ground state on an $L = 10$
chain. The top panel corresponds to the fully dimerized state $\ket{\Phi_{\textrm{dimer}}}$
\eqref{eq: dimer state}; the middle and bottom panels follow the evolution as $t_2$ increases
to the critical point $t_1 = t_2$. In all cases the block-diagonal dipole structure~\eqref{eq: RDM direct sum}
is well established by $g \sim 2$, confirming that the asymptotic limit is reached at
$g \sim \mathrm{O}(1)$ for chain sizes $L \sim 4$--$14$.
The saturation of the entanglement entropy to its ultrastrong-coupling 
asymptote at $g \sim \mathcal{O}(1)$, visible in Fig.~\ref{fig: LM entanglement 
as function of g and t's}, is independent of system size. This follows directly 
from the structure of the reduced density matrix~\eqref{eq: photon DM state 0}, indicating that 
the crossover is controlled by the condition $g^2 (\Delta\mathcal{P})^2 \sim 1$, 
where $\Delta\mathcal{P}$ is the dipole difference between neighboring sectors. 
Since the minimum nonzero dipole difference between adjacent sectors is 
$\Delta\mathcal{P} = 1$ (set by the single-site contribution to $\mathcal{P}$, 
independently of $L$) the Gaussian suppression of inter-sector coherences 
is effective at $g \sim \mathcal{O}(1)$ regardless of system size.

\begin{figure*}[t]
\centering
\includegraphics[width=\textwidth]{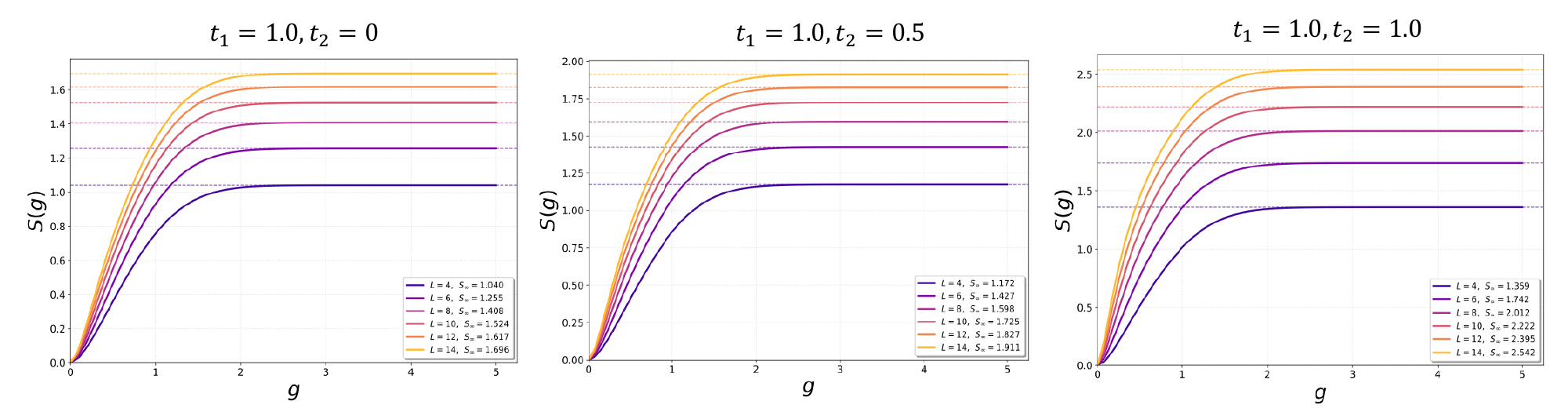}
\caption{Light-matter entanglement entropy for
$\ket{\Psi_0(g)} = \mathcal{U}(g)\ket{\Phi}\otimes\ket{0}$, where $\ket{\Phi}$ is the SSH
ground state, as a function of $g$ and $L$ for the completely dimerized chain (top),
intermediately dimerized chain (middle), and critical chain (bottom). Solid curves: exact
entropy from~\eqref{eq: photon DM state 0}. Dotted lines: ultrastrong-coupling
asymptote~\eqref{eq: RDM direct sum}. In all cases the entropy saturates to the asymptotic
value at $g \sim 2$.}
\label{fig: LM entanglement as function of g and t's}
\end{figure*}

Figure~\ref{fig: LM entanglement as function of g and t's} plots the entanglement entropy
as a function of $g$ for several system sizes. Solid curves show the exact entropy extracted
from~\eqref{eq: photon DM state 0}; dotted lines show the ultrastrong-coupling
asymptote from~\eqref{eq: RDM direct sum}. The saturation to the asymptotic value at
$g \sim \mathrm{O}(1)$ is consistent across all three SSH parameter regimes shown.

\noindent\textit{Ultrastrong coupling and logarithmic scaling.---}
To characterize the system-size scaling, we analyze the asymptotic density
matrix~\eqref{eq: RDM direct sum} obtained by tracing out the photon from
$\ket{\Psi_0(g\to\infty)} = \mathcal{U}(g\to\infty)\ket{\Phi_{\textrm{dimer}}}\otimes\ket{0}$.
In this limit the Schmidt decomposition is exact:
\begin{equation}
\label{eq: photon reduced DM at strong coupling}
\rho^{\textrm{dimer}}_{\infty}
=
\sum^{L/2}_{k=0} p_k\,\ket{\Omega_k}\bra{\Omega_k}\,,
\quad
p_k = 2^{-L/2}\begin{pmatrix} L/2 \\ k \end{pmatrix}\,.
\end{equation}
The Schmidt eigenvalues $\{p_k\}$ are precisely the dipole-sector weights, and the entanglement
entropy~\eqref{eq: dipole expression for EE} becomes
\begin{equation}
\label{eq: photon EE at strong coupling}
S^{\textrm{dimer}}_{\infty}
= -\sum^{L/2}_{k=0} p_k\,\log p_k
\approx -\int d\mathcal{P}\,p(\mathcal{P})\,\log p(\mathcal{P})\,,
\end{equation}
where the integral form follows from the large-$L$ Gaussian
limit~\eqref{eq: Schmidt eigenvalues dimer state exponential form}.
Evaluating the Gaussian integral yields the logarithmic scaling law
\begin{equation}
\label{eq: S scaling for dimer state}
S^{\textrm{dimer}}_{\infty} \approx \tfrac{1}{2}\log L + C\,.
\end{equation}

The origin of this scaling is transparent: in the large-$L$ limit the distribution~\eqref{eq: Schmidt eigenvalues dimer state exponential form} has $\mathrm{O}((L/4)^{1/2})$ appreciable eigenvalues each of order $\mathrm{O}((L/4)^{-1/2})$, so that $S_\infty \sim -L^{1/2}\times(L^{-1/2}\log L^{-1/2}) \sim \log L$. The photon resolves only a single collective coordinate $\mathcal{P}$ whose fluctuations grow as $L^{1/2}$; the entanglement is therefore set by the statistics of this collective variable, not by the full many-body Hilbert space. In this sense, correlations within each dipole sector are left entirely intact by the photonic environment, and this selectivity is responsible for keeping the entanglement growth subextensive.

\begin{figure}
\includegraphics[width=8.0cm]{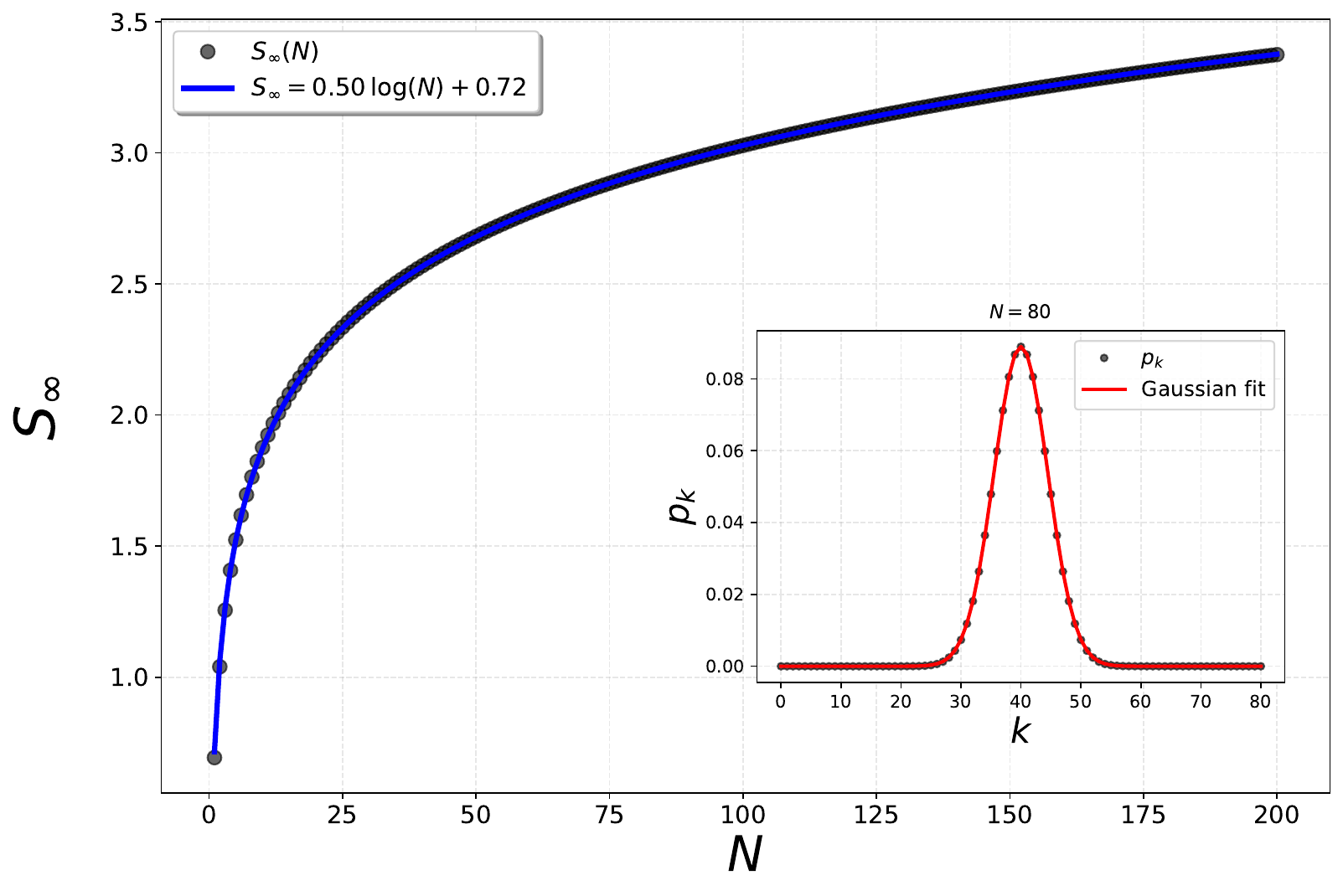}
\caption{Scaling of the light-matter entanglement entropy $S_\infty$~\eqref{eq: photon EE at strong coupling}
as a function of $N = L/2$. The solid blue line is a logarithmic fit consistent
with~\eqref{eq: S scaling for dimer state}.
Inset: Schmidt eigenvalues $p_k$ for $N = 80$ ($L = 160$), well described by a Gaussian
centered at $k = N/2$, confirming Eq.~\eqref{eq: Schmidt eigenvalues dimer state exponential form}.}
\label{fig: Entanglement Scaling - Dimer}
\end{figure}

Figure~\ref{fig: Entanglement Scaling - Dimer} shows $S^{\textrm{dimer}}_\infty$ as a function
of $N = L/2$. The logarithmic fit~\eqref{eq: S scaling for dimer state} describes the numerical
data over two decades in $N$. The inset plots the Schmidt eigenvalues $p_k$ for $N = 80$
($L = 160$), confirming the Gaussian form~\eqref{eq: Schmidt eigenvalues dimer state exponential form}
centered at $k = N/2$. 

\noindent\textit{Dependence on coupling strength.---}
The logarithmic scaling~\eqref{eq: S scaling for dimer state} holds at ultrastrong coupling,
but the coefficient of $\log L$ evolves continuously with $g$. More generally, whenever the
Schmidt eigenvalue distribution is Gaussian with a power-law variance $\sigma(L)^2 \sim L^\alpha$,
\begin{equation}
\label{eq: distribution of p as function of dipole - general}
p(\mathcal{P}) \approx (2\pi\sigma(L)^2)^{-1/2}\,
e^{-\frac{\mathcal{P}^2}{2\,\sigma(L)^{2}}}\,,
\quad \sigma(L)^2 \sim L^\alpha\,,
\end{equation}
the entanglement entropy~\eqref{eq: dipole expression for EE} reduces to
\begin{equation}
\label{eq: Gaussian EE}
\begin{split}
S_{\infty}
&\,= -\Big\langle\log\Big[(\pi\sigma(L))^{-1/2}
e^{-\frac{\mathcal{P}^2}{\sigma(L)}}\Big]\Big\rangle \\
&\,= \tfrac{1}{2}\log(2\pi e\,\sigma(L)^2) \\
&\,\sim \tfrac{\alpha}{2}\log L\,,
\end{split}
\end{equation}
where $\langle\cdot\rangle$ denotes the average over~\eqref{eq: distribution of p as function of dipole - general}
and $\alpha$ is the scaling exponent of the variance. The logarithmic scaling is therefore
robust: it persists for any matter state whose dipole distribution has a power-law spread, with
the coefficient $\alpha/2$ encoding the rate at which dipole fluctuations grow with system size.

\begin{figure}
\includegraphics[width=9.0cm]{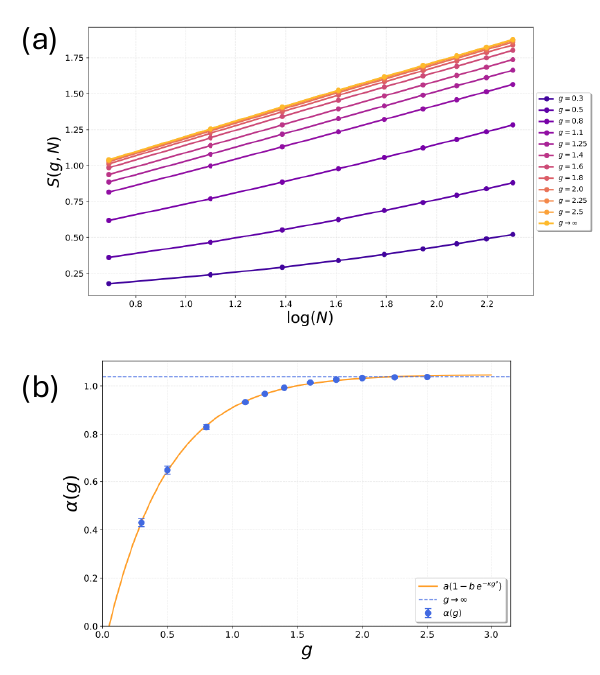}
\caption{Coupling-strength dependence of the entanglement scaling for the dimerized
state $(t_1,t_2) = (1,0)$. (a) Entanglement entropy vs $\log N$ for several values of $g$,
showing good agreement with $S = \alpha(g)\log N$ across the full crossover.
(b) Fitted coefficient $\alpha(g)$ as a function of $g$, well described by a stretched
exponential that saturates to the ultrastrong-coupling value $\alpha(g\to\infty)$.}
\label{fig: S vs Log N and alpha (g) for dimerized state}
\end{figure}

\begin{figure}
\includegraphics[width=9.0cm]{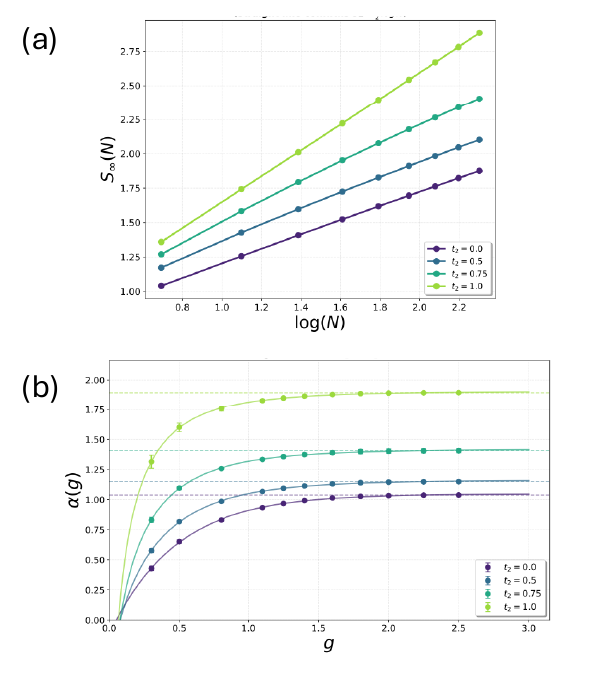}
\caption{Entanglement scaling away from the dimerized limit.
(a) Ultrastrong-coupling entropy $S_\infty$ vs $\log N$ for several values of $t_2$.
All curves are consistent with logarithmic scaling; the prefactor grows with the
correlation length of the SSH chain and is largest at the critical point $t_2 = t_1$.
(b) Coefficient $\alpha(g)$ as a function of $g$ for the same values of $t_2$,
showing that $\alpha$ saturates to the strong-coupling value in all cases.}
\label{fig: Entaglement-varying_over_t2}
\end{figure}

Figures~\ref{fig: S vs Log N and alpha (g) for dimerized state} and
\ref{fig: Entaglement-varying_over_t2} substantiate this picture numerically.
Figure~\ref{fig: S vs Log N and alpha (g) for dimerized state}(a) shows $S(g,N)$ vs $\log N$
for the dimerized chain, confirming the form $S = \alpha(g)\log N$ across the full range
of $g$. Panel (b) shows that the coupling-dependent coefficient $\alpha(g)$ increases
monotonically from zero and saturates at its ultrastrong-coupling value, well fitted by a
stretched exponential. Figure~\ref{fig: Entaglement-varying_over_t2} extends this analysis
to varying $t_2$: panel (a) shows that the logarithmic scaling persists for all values of $t_2$,
with the prefactor $\alpha$ growing with the correlation length of the SSH chain and reaching
its maximum at the critical point $t_2 = t_1$; panel (b) confirms the same monotonic
saturation of $\alpha(g)$ independently of $t_2$.

\subsection{Photo-induced matter entanglement}
\label{sec: Photo-induced matter entanglement}

Having characterized the light-matter entanglement, we now turn to the entanglement
within the matter sector itself. The question is how the photon-mediated long-range
correlations, imprinted by the PZW circuit onto the matter state, modify the spatial
entanglement scaling.

We find that in the ultrastrong coupling limit, the dominant eigenstates of $\rho_\infty$
carry a half-chain entanglement entropy $S^{\mathrm{matter}}_\infty \sim \log\ell$, where
$\ell$ is the subsystem size. This is the same functional form as the light-matter
entanglement~\eqref{eq: S scaling for dimer state}, and its origin can be traced to the
same mechanism: the photon resolves the dipole of the full chain, which in turn constrains
the distribution of dipole fluctuations between any bipartition.

To make this precise, we again work in the perfectly dimerized limit ($t_2 = 0$) and follow
the analysis of Sec.~\ref{sec: Light-matter entanglement}. According to
Eq.~\eqref{eq: Schmidt eigenvalues dimer state exponential form}, the distribution of
Schmidt eigenvalues of $\rho^{\textrm{dimer}}_\infty$ is dominated by the typical
sector $\ket{\Omega_k}$ with $k \approx N/2$. To compute the spatial entanglement entropy
imprinted onto the matter by the photon, we introduce a bipartition through the chain midpoint,
dividing it into regions $A$ and $B$ each supporting $N/2$ dimers ($\ell = N/2$). In the large
$N$ limit the typical state $\ket{\Omega_{k=N/2}}$ contains $N/2$ broken dimers distributed
between $A$ and $B$: let $s$ and $N/2 - s$ denote the number of broken dimers in $A$ and $B$
respectively, with $s = 0,\ldots,N/2$.

With $d_A(s) = \binom{N/2}{s}$ configurations in $A$ and
$d_B(s) = \binom{N/2}{N/2-s}$ configurations in $B$, the state $\ket{\Omega_{k=N/2}}$
is spanned by
$
\Big\{
\ket{\alpha_s,s}\otimes\ket{\beta_s,N/2-s},\quad
\alpha_s = 0,\ldots,d_A(s)-1\,,\quad
\beta_s = 0,\ldots,d_B(s)-1
\Big\}\,,
$
where $\alpha_s$ and $\beta_s$ label distinct dimer configurations in each partition,
and the dimer states form an orthonormal set. The normalized state reads
\begin{equation}
\ket{\Omega_{k=N/2}}
=
\begin{pmatrix} N \\ N/2 \end{pmatrix}^{-1/2}
\sum_{s}
\sum_{\alpha_s}
\sum_{\beta_s}
\ket{\alpha_s,s}\otimes\ket{\beta_s,N/2-s}\,.
\end{equation}

Defining the normalized subsystem states
\begin{equation}
\ket{A^{N/2}_s} \equiv
\begin{pmatrix} N/2 \\ s \end{pmatrix}^{-1/2}
\sum^{d_A(s)-1}_{\alpha_s=0} \ket{\alpha_s,s}
\end{equation}
and
\begin{equation}
\ket{B^{N/2}_s} \equiv
\begin{pmatrix} N/2 \\ N/2-s \end{pmatrix}^{-1/2}
\sum^{d_B(s)-1}_{\beta=0} \ket{\beta,N/2-s}\,,
\end{equation}
the state $\ket{\Omega_{k=N/2}}$ takes the Schmidt form
\begin{subequations}
\begin{equation}
\ket{\Omega_{k=N/2}}
= \sum^{N/2}_{s=0} \sqrt{p_s}\,\ket{A^{N/2}_s}\otimes\ket{B^{N/2}_s}\,,
\end{equation}
\begin{equation}
\label{eq: hypergeometric distribution}
p_s =
\frac{
\begin{pmatrix} N/2 \\ s \end{pmatrix}
\begin{pmatrix} N/2 \\ N/2-s \end{pmatrix}
}{
\begin{pmatrix} N \\ N/2 \end{pmatrix}
}\,.
\end{equation}
\end{subequations}
The Schmidt eigenvalues~\eqref{eq: hypergeometric distribution} follow a hypergeometric
distribution, which in the large-$N$ limit converges to a Gaussian,
\begin{equation}
\label{eq: Gaussian limit of hypergeometric distribution}
p_s = \frac{1}{\sqrt{2\pi\sigma^2}}\,
e^{-\frac{(s-\mu)^2}{2\sigma^2}}\,,
\quad \mu = N/4\,,\quad \sigma^2 = N/16\,.
\end{equation}
Applying Eq.~\eqref{eq: Gaussian EE} with $\sigma^2 = N/16 \sim \ell^1$, the half-chain
entanglement entropy of the typical state is
\begin{equation}
\label{eq: half-chain EE}
S^{\mathrm{matter}}_\infty \sim \tfrac{1}{2}\log(N/2) \sim \tfrac{1}{2}\log\ell\,.
\end{equation}

Equation~\eqref{eq: half-chain EE} reveals a notable feature of the PZW circuit: at
ultrastrong coupling, both the light-matter entanglement~\eqref{eq: S scaling for dimer state}
and the spatial entanglement of the photon-dressed matter state scale logarithmically with
system size. This is notable because the states~\eqref{eq: general state as fcn of m} are
non-critical; rather, they are obtained by a unitary transformation from a gapped, disentangled
product state, yet they carry an entanglement signature ordinarily associated with critical
systems. The mechanism is distinct from the conformal-field-theory origin of logarithmic
scaling in critical chains~\cite{holzhey1994geometric, calabrese2004entanglement}: here,
the logarithm arises because the cavity photon couples globally to the many-body dipole,
restructuring the matter entanglement through long-range correlations rather than through
local criticality.

\section{Summary and Discussion}
\label{sec: Summary and Discussion}

We have presented an analytical framework for characterizing the entanglement structure of cavity materials, wherein 
a cavity photon mode interacts with an extended one-dimensional system. 
We have introduced a quantum circuit that generates light-matter entanglement via Power-Zienau-Woolley (PZW) unitary transformation, whose generator involves the coupling of the {photon's position quadrature $X \sim a + a^{\dagger}$} with the many-body dipole operator $\mathcal{P}$. Leveraging the Gaussian nature of the photon mode, we have derived a closed-form expression for the reduced density matrix obtained by tracing out the
photon from the PZW-circuit state, valid at all values of the coupling strength. 
The Gaussian suppression of off-diagonal matrix elements between matter states of unequal dipole encodes the full crossover from weak to ultrastrong coupling in a single exact expression, yielding a comprehensive description for the behavior of entanglement in the hybrid light-matter systems as a function of light-matter coupling and system size.

At weak coupling, perturbative corrections of the reduced density matrix due to the light-matter interaction take a Lindbladian form with the many-body
dipole $\mathcal{P}$ as the jump operator. The Hermitian nature of the jump operator
reflects the fact that the photon acts as a monitor of the dipole: it acquires
information about dipole fluctuations while leaving states of definite dipole unaffected.
As a result, the entanglement entropy is controlled entirely by the dipole variance, and
grows quasi-linearly with system size for a half-filled chain.

At ultrastrong coupling, the Gaussian suppression is complete and the reduced density
matrix becomes exactly block-diagonal in dipole sectors, reflecting an emergent dipole
symmetry dynamically imposed by the photon field. The entanglement entropy reduces to the
Shannon entropy of the dipole-sector weight distribution, and scales logarithmically with
system size. 

We have applied the general considerations for the PZW circuit in the context where a single cavity mode entangles with a one-dimensional half-filled chain with nearest-neighbor hopping amplitudes $t_1$ (intra unit cell) and $t_2$ (inter unit cell). This system permits us to investigate the entanglement properties both as a function of the light-matter coupling, as well as of the correlation length by varying the hopping amplitudes - from a trivial insulator in the complete dimerized limit $(t_2/t_1=0)$, all the way to the metallic state $(t_2/t_1=1)$.
For the half-filled SSH chain, both the light-matter entanglement and the
half-chain spatial entanglement of the photon-dressed matter state grow as $(\alpha/2)\log
L$, sharing the same mechanism: the photon resolves only the single collective coordinate
$\mathcal{P}$, whose fluctuations grow as a power of $L$, so the entanglement is
controlled by the statistics of this variable rather than by the full many-body Hilbert space. 
In particular, in the fully dimerized limit $t_2= 0$, we have explored the exact representation of the ground state to analytically derive the dipole distribution and the logarithmic scaling of the entanglement entropy, providing insight into the complementary numerical analysis carried out for other parameter regimes. Numerically, the ultrastrong-coupling asymptote is already well established at $g \sim \mathrm{O}(1)$.

The prefactor $\alpha/2$ is not universal in the CFT sense: it evolves continuously with
$g$, saturating at its ultrastrong-coupling value, and grows with the correlation length
of the SSH chain, reaching its maximum at the critical point $t_1 = t_2$. Both
logarithmic scalings arise in states that are topologically trivial, obtained by a
unitary transformation from a gapped product state, and are therefore qualitatively
distinct in origin from the logarithmic entanglement of critical one-dimensional
systems~\cite{holzhey1994geometric, calabrese2004entanglement}. In critical systems the
coefficient of $\log L$ is fixed by the central charge of the underlying CFT. Here, the
logarithm is a direct consequence of non-local photon-mediated coupling rather than
conformal invariance, and the prefactor encodes the rate at which dipole fluctuations
grow with system size.

The emergent dipole symmetry at ultrastrong coupling has a notable consequence for matter
observables: operators that do not conserve $\mathcal{P}$ acquire vanishing expectation
values, suppressing single-particle hopping while leaving dipole-conserving correlated
processes active. This structure is remarkably analogous to the dipole-conserving
Hamiltonians that underlie Hilbert space
fragmentation~\cite{sala2020ergodicity, khemani2020localization, moudgalya2022review},
where conservation of the dipole moment partitions the Hilbert space into dynamically
disconnected sectors. A key distinction is that in our model the dipole constraint is not
intrinsic to the matter Hamiltonian but is dynamically generated by the cavity photon
field. Whether this photon-induced constraint can drive a genuine fragmentation of the
effective matter Hilbert space, and what its dynamical signatures would be in
cavity-matter systems, are interesting open questions.

Our results also provide analytical insight into the subextensive entanglement scaling
observed numerically in Ref.~\onlinecite{shaffer2024entanglement} for one-dimensional chains
in a cavity. The mechanism identified here, namely, the photon resolving a single collective
coordinate whose fluctuations grow subextensively, offers a mechanism
for why cavity-induced long-range interactions do not produce volume-law entanglement:
the mode-uniform nature of the photon coupling selects a single degree of freedom, and
the entanglement is bounded by the information content of that coordinate rather than by
the full many-body Hilbert space.

We conclude by mentioning possible extensions of this work. 
Besides the suggestive connections between the emergent dipole symmetry at strong coupling
and Hilbert space fragmentation mentioned before, which warrant a separate investigation,
generalizations to multi-mode cavities, higher-dimensional matter systems,
or interacting fermionic chains could inform new entanglement patterns beyond those presented in this context.
In particular, motivated by experimental signatures linked to disturbances caused by cavity fields on the topological properties of quantum Hall systems \cite{Appugliese_Breakdown2022}, an extension of this analysis to two-dimensional topological systems, such as fractional Chern insulators both in lattice systems  \cite{neupert2011fractional,Sun-2011,Sheng-2011, Tang-2011, Regnault2011}, as well as in moir\'e materials \cite{cao2025fractional, lu2024fractional, lu2025electromagnetic}, is an outstanding question.

Moreover, the PZW circuit invites a possibly rich connection with the physics of measurement-induced phase transitions
(MIPTs)~\cite{li2018quantum, skinner2019measurement, chan2019unitary}. In MIPT
settings, competing unitary dynamics and measurements drive a transition in the
entanglement scaling of quantum trajectories, typically between a volume-law phase
and an area-law phase. 
The strong coupling limit of the light-matter circuit can be interpreted as ``long time" regime of a unitary dynamical evolution, which is described by a new entanglement scaling regime. It would be fundamental to explore the competition between this unitary dynamics and measurement channels, and whether this competition could give rise to new classes of measure-induced phase transitions.
\\

\noindent
\textit{Acknowledgments.}
The author thanks Martin Claassen, Daniel Shaffer, and Ajit Srivastava for an earlier collaboration, Ref.~\onlinecite{shaffer2024entanglement}, which motivated this work, and Tianhong Lu for assistance in preparing the cavity quantum matter setup figure.
This research is supported by the National Science Foundation under CAREER Award No.~DMR-2441621. 
Computational resources provided by the Indiana Jestream2 Cluster via NSF ACCESS are gratefully acknowledged. This work was performed in part at the Aspen Center for Physics, which is supported by National Science Foundation grant PHY-2210452. Generous support from the Ramond Family Fund through the Aspen Center for Physics is also gratefully acknowledged.

\bibliographystyle{apsrev4-1}
\bibliography{bibliography}

\end{document}